# Message randomization and strong security in quantum stabilizer-based secret sharing for classical secrets

**Ryutaroh Matsumoto[1,2]**

## Abstract

We improve the flexibility in designing access structures of quantum stabilizer-based secret sharing schemes for classical secrets, by introducing message randomization in their encoding procedures. We generalize the Gilbert–Varshamov bound for deterministic encoding to randomized encoding of classical secrets. We also provide an explicit example of a ramp secret sharing scheme with which multiple symbols in its classical secret are revealed to an intermediate set, and justify the necessity of incorporating strong security criterion of conventional secret sharing. Finally, we propose an explicit construction of strongly secure ramp secret sharing scheme by quantum stabilizers, which can support twice as large classical secrets as the McEliece–Sarwate strongly secure ramp secret sharing scheme of the same share size and the access structure.

## 1 Introduction

Secret sharing is a scheme to share a secret among multiple participants so that only *qualified* sets of participants can reconstruct the secret, while *forbidden* sets have no information about the secret [14,49,53]. A piece of information received by a participant is called a *share*. A

This is one of several papers published in *Designs, Codes and Cryptography* comprising the "Special Issue on Coding and Cryptography 2019".

Orally presented at the unrefereed poster session in 2019 IEEE ISIT, Paris, France. This research is partly supported by JSPS Grant No. 17K06419.

✉ Ryutaroh Matsumoto
ryutaroh@ict.e.titech.ac.jp

1    Department of Information and Communications Engineering, Tokyo Institute of Technology, Tokyo, Japan

2    Department of Mathematical Sciences, Aalborg University, Ålborg, Denmark

set of participants that is neither qualified nor forbidden is said to be *intermediate*. If there is no intermediate set, a secret sharing scheme is said to be *perfect*, otherwise said to be *ramp* [5,54]. There is an upper bound on the size of secret for fixed size of shares, when secret sharing is perfect. On the other hand, the size of secret can be arbitrarily large for fixed size of shares in ramp schemes. In this paper we consider ramp schemes, in other words, we allow intermediate sets of participants or shares.

Both secret and shares are traditionally classical information. There exists a close connection between secret sharing and classical error-correcting codes [4,12,15,16,29,34,44].

After the importance of quantum information became well-recognized, secret sharing schemes with quantum shares were proposed [13,20,23,25,51]. A connection between quantum secret sharing and quantum error-correcting codes has been well-known for many years [13,18,20,32,33,36,37,48,51]. Well-known classes of quantum error-correcting codes are the CSS codes [11,52], the stabilizer codes [9,10,19] and their nonbinary generalizations [3,26,42].

The access structure of a secret sharing scheme is the set of qualified sets, that of intermediate sets and that of forbidden sets. When both secret and shares are classical information, encoding of secrets to shares are almost always randomized, that is, for a fixed secret, shares are randomly chosen from a set determined by the secret [14,49,53]. By message randomization we mean this kind of randomized encoding of secrets to shares. It was shown that some randomness in encoders is indispensable with classical shares [6–8].

In contrast with classical shares, Gottesman [20, Theorem 3] proved that message randomization does not offer any advantage when both secret and shares are quantum information, and that use of unitary encoding of quantum secret to quantum shares is sufficient. Probably because of Gottesman's observation, secret sharing schemes based on quantum error-correcting codes have not used message randomization, as far as this author knows.

In our previous research [38,40], we expressed secret sharing for classical secrets based on quantum stabilizer codes by linear codes, and expressed qualified and forbidden sets in terms of the linear codes associated with quantum stabilizers. By using that, we gave a Gilbert-Varshamov-type existence condition of secret sharing schemes with given parameters, and proved that there exist infinitely many access structures that can be realized by quantum stabilizer codes but cannot be realized by any classical information processing.

However, there are some drawbacks in our proposal [38,40]. For example, any $n - 1$ participants out of $n$ participants can be made forbidden, for example, by Shamir's scheme. But such an access structure cannot be realized by [38,40]. The first goal of this paper is to make the stabilizer-based secret sharing more flexible in designing access structures by introducing message randomization in the encoding. In our previous proposal [38,40], shares are deterministic functions of secrets. The proposed scheme in this paper includes [38,40] as a special case.

Ordinary ramp schemes have the following security risk: Suppose that classical secret is $\mathbf{m} = (m_1, \ldots, m_k)$, and an intermediate set has $\ell(\geq 1)$ symbol of information about $\mathbf{m}$. Then that intermediate set sometimes knows $m_i$ explicitly for some $i$. This insecurity was mentioned in [44,54]. Iwamoto and Yamamoto [24] explicitly constructed such an example with classical secret and classical shares, and Zhang and Matsumoto [55] did with quantum shares. In order to address this security risk, Yamamoto [54] introduced the notion of strong security into ramp schemes: A secret sharing scheme with classical secret $\mathbf{m} = (m_1, \ldots, m_k)$ is said to be *strongly secure* [24] if any $(k - \ell)$ symbols in $\mathbf{m}$ is always statistically independent of shares in an intermediate set that has $\ell$ symbol of information about $\mathbf{m}$, for $\ell = 1, \ldots, k-1$. Recently Martínez-Peñas constructed communication efficient and strongly secure ramp schemes with classical shares [35]. The second goal of this paper is to give an explicit construction of strongly secure ramp secret sharing for classical secrets based on quantum stabilizer codes, by extending the previous construction [38,40].

Strong security concerns with secrecy of parts of a message. The secrecy of parts of a message has also been studied for network coding [21,28,41,50] and wiretap channel coding [22,27].

Note that, throughout in this paper, secrets are classical and shares are quantum. The value added by randomization does not contradict with Gottesman's observation [20, Theorem 3] that focused on quantum secrets.

This paper is organized as follows: Sect. 2 introduces necessary notations and proposes randomized encoding for quantum stabilizer-based secret sharing. Section 3 clarifies the access structure of the proposed scheme. Section 4 analyzes the amount of information leaked to an intermediate set, which will be used for the strong security later. Section 5 generalizes the Gilbert-Varshamov existential condition for secret sharing schemes from one given in [38,40]. Section 6 introduces a strong security criterion and an explicit construction with strong security based on Reed-Solomon codes. Then we compare the proposed construction with the McEliece-Sarwate strongly secure ramp secret sharing scheme [44].

## 2 Randomized encoding and its access structures

### 2.1 Preliminaries

Let $A \subset \{1, \ldots, n\}$ be a set of shares (or equivalently participants), $\overline{A} = \{1, \ldots, n\} \setminus A$, and $\text{Tr}_{\overline{A}}$ the partial trace over $\overline{A}$. For a density matrix $\rho$, $\text{col}(\rho)$ denotes its column space. When $\text{col}(\rho_1), \ldots, \text{col}(\rho_n)$ are orthogonal to each other, that is, $\rho_i \rho_j = 0$ for $i \neq j$, we can distinguish $\rho_1, \ldots, \rho_n$ by a suitable projective measurement with probability 1. Since density matrices are quantum generalization of probability distributions [45], the quantum randomized encoding of a secret can be expressed as a density matrix.

**Definition 1** [38,40] Let $\rho_A(\mathbf{m})$ be the density matrix of shares in $A$ encoded from a classical secret $\mathbf{m}$. We say $A$ to be qualified if $\text{col}(\rho_A(\mathbf{m})))$ and $\text{col}(\rho_A(\mathbf{m}'))$ are orthogonal to each other for different classical secrets $\mathbf{m}, \mathbf{m}'$. We say $A$ to be forbidden if $\rho_A(\mathbf{m})$ is the same density matrix regardless of classical secret $\mathbf{m}$. By an access structure we mean the set of qualified sets and the set of forbidden sets.

Let $p$ be a prime number, $\mathbf{F}_p$ the finite field with $p$ elements, and $\mathbf{C}_p$ the $p$-dimensional complex linear space. The quantum state space of $n$ qudits is denoted by $\mathbf{C}_p^{\otimes n}$ with its orthonormal basis $\{|\mathbf{v}\rangle : \mathbf{v} \in \mathbf{F}_p^n\}$.

For two vectors $\mathbf{a}, \mathbf{b} \in \mathbf{F}_p^n$, denote by $\langle \mathbf{a}, \mathbf{b} \rangle_E$ the standard Euclidean inner product. For two vectors $(\mathbf{a}|\mathbf{b})$ and $(\mathbf{a}'|\mathbf{b}') \in \mathbf{F}_p^{2n}$, we define the standard symplectic inner product

$$\langle (\mathbf{a}|\mathbf{b}), (\mathbf{a}'|\mathbf{b}') \rangle_s = \langle \mathbf{a}, \mathbf{b}' \rangle_E - \langle \mathbf{a}', \mathbf{b} \rangle_E.$$

For an $\mathbf{F}_p$-linear space $C_S \subset \mathbf{F}_p^{2n}$, $C_S^{\perp_s}$ denotes its orthogonal space in $\mathbf{F}_p^{2n}$ with respect to $\langle \cdot, \cdot \rangle_s$. Throughout this paper we always assume $\dim C_S = n - k - s$ and $C_S \subseteq C_S^{\perp_s}$. We will use $k$ to denote the number of symbols in classical secrets and $s(\geq 0)$ to denote amount of randomness in encoding. We also assume that we have $C_S \subset C_R \subseteq C_R^{\perp_s}$ and $\dim C_R = n - s$.

Let $X$ be the $p \times p$ complex unitary matrix defined by $X|i\rangle = |i + 1\rangle$ for $i \in \mathbf{F}_p$, and $Z$ the $p \times p$ complex unitary matrix defined by $Z|i\rangle = \omega^i |i\rangle$, where $\omega$ is a primitive $p$-th root of 1 in the complex numbers. For $(\mathbf{a}|\mathbf{b}) = (a_1, \ldots, a_n|b_1, \ldots, b_n) \in \mathbf{F}_p^{2n}$, define the $p^n \times p^n$ complex unitary matrix $X(\mathbf{a})Z(\mathbf{b}) = X^{a_1}Z^{b_1} \otimes \cdots \otimes X^{a_n}Z^{b_n}$ as defined in [26]. An $[[n, k+s]]_p$ quantum stabilizer codes $Q$ encoding $k+s$ qudits into $n$ qudits can be defined as a simultaneous eigenspace of all $X(\mathbf{a})Z(\mathbf{b})$ $((\mathbf{a}|\mathbf{b}) \in C_{\mathrm{S}})$. Unlike [26] we do not require the eigenvalue of $Q$ to be one, which means that the eigenvalue of $|\varphi\rangle \in Q$ is not required to be one for $X(\mathbf{a})Z(\mathbf{b})$ $((\mathbf{a}|\mathbf{b}) \in C_{\mathrm{S}})$.

## 2.2 Proposed randomized encoding

Witt's Lemma [1], [2, Chapter 7] states that if there exist two subspaces $V_1, V_2 \subset W$ and their bijective linear map $\iota : V_1 \to V_2$ preserving symplectic inner products in $V_1$, then there always exists a symplectic isometry $\kappa : W \to W$, that is, a bijective linear map preserving the symplectic inner products in $W$, such that the restriction of $\kappa$ to $V_1$ is $\iota$. Let $W = \mathbf{F}_p^{2n}$, $V_1 = C_{\mathrm{R}}$, $V_2 = \{(a_1, \ldots, a_{n-s}, 0, \ldots, 0|0, \ldots, 0)|a_i \in \mathbf{F}_p, i = 1, \ldots, n-s\}$, and $V_{\max} = \{(a_1, \ldots, a_n|0, \ldots, 0)|a_i \in \mathbf{F}_p, i = 1, \ldots, n\}$. Then there exists a bijective linear map $\iota : V_1 \to V_2$ satisfying the assumptions in Witt's Lemma and we also have $V_{\max}^{\perp s} = V_{\max}$. Let $\kappa : W \to W$ as implied by Witt's Lemma, and $C_{\max} = \kappa^{-1}(V_{\max})$. We have $C_{\mathrm{S}} \subseteq C_{\mathrm{R}} \subseteq C_{\max} \subseteq C_{\mathrm{R}}^{\perp s} \subseteq C_{\mathrm{S}}^{\perp s}$ with $C_{\max} = C_{\max}^{\perp s}$. Note that $C_{\max}$ is not unique and usually there are many possible choices of $C_{\max}$. We have $\dim C_{\max} = n$ and have an isomorphism $f : \mathbf{F}_p^k \to C_{\mathrm{S}}^{\perp s}/C_{\mathrm{R}}^{\perp s}$ as linear spaces without inner products. Since $C_{\max} = C_{\max}^{\perp s}$, $C_{\max}$ defines an $[[n, 0]]_p$ quantum stabilizer code $Q_0$. Without loss of generality we may assume $Q_0 \subset Q$. Let $|\varphi\rangle \in Q_0$ be a quantum state vector. Since $C_{\max} = C_{\max}^{\perp s}$, for a coset $V \in C_{\mathrm{S}}^{\perp s}/C_{\max}$ and $(\mathbf{a}|\mathbf{b}), (\mathbf{a}'|\mathbf{b}') \in V$, $X(\mathbf{a})Z(\mathbf{b})|\varphi\rangle$ and $X(\mathbf{a}')Z(\mathbf{b}')|\varphi\rangle$ differ by a constant multiple in $\mathbf{C}$ and physically express the same quantum state in $Q$. By an abuse of notation, for a coset $V \in C_{\mathrm{S}}^{\perp s}/C_{\max}$ we will write $|V\varphi\rangle$ to mean $X(\mathbf{a})Z(\mathbf{b})|\varphi\rangle$ $((\mathbf{a}|\mathbf{b}) \in V)$.

For a given classical secret $\mathbf{m} \in \mathbf{F}_p^k$, we consider the following secret sharing scheme with $n$ participants:

1. $f(\mathbf{m})$ is a coset of $C_{\mathrm{S}}^{\perp s}/C_{\mathrm{R}}^{\perp s}$ and $f(\mathbf{m})$ can also seen as a subset of $C_{\mathrm{S}}^{\perp s}/C_{\max}$. Choose $V \in f(\mathbf{m}) \subset C_{\mathrm{S}}^{\perp s}/C_{\max}$ uniformly at random. Prepare the quantum codeword $|V\varphi\rangle \in Q$ that corresponds to the classical secret $\mathbf{m}$.
2. Distribute each qudit in the quantum codeword $|V\varphi\rangle$ to a participant.

Since there are $p^s$ choices of $V$ above, the density matrix of $n$ shares is

$$\rho(\mathbf{m}) = \frac{1}{p^s} \sum_{V \in f(\mathbf{m})} |V\varphi\rangle\langle V\varphi|.$$

**Remark 2** The encoding procedure in [38,40] corresponds to the special case $C_{\mathrm{R}} = C_{\max} = C_{\mathrm{R}}^{\perp s}$ and $s = 0$ in the above proposed scheme.

**Example 3** Let $p = 3$, $n = 4$, $k = s = 2$. A basis of the doubly-extended $[4, 2, 3]_3$ Reed-Solomon code over $\mathbf{F}_3$ consists of

$$\mathbf{v}_1 = (1, 1, 1, 0),$$
$$\mathbf{v}_2 = (0, 1, 2, 1).$$

By using them, we define $C_{\mathrm{S}} = \{\mathbf{0}\}$, $C_{\mathrm{R}}$ as the linear space spanned by $\{(\mathbf{v}_1|\mathbf{0}), (\mathbf{0}|\mathbf{v}_1)\}$, and $C_{\max}$ as the linear space spanned by $\{(\mathbf{v}_1|\mathbf{0}), (\mathbf{v}_2|\mathbf{0}), (\mathbf{0}|\mathbf{v}_1), (\mathbf{0}|\mathbf{v}_2)\}$. Let

$$\mathbf{v}_3 = (0, 1, 1, 0).$$

Then $C_R^{\perp_s}$ is spanned by $C_{\max} \cup \{(\mathbf{v}_3|\mathbf{0}), (\mathbf{0}|\mathbf{v}_3)\}$. Let

$$\mathbf{v}_4 = (0, 0, 0, 1).$$

$C_S^{\perp_s} = \mathbf{F}_3^8$ and we can use $\{(\mathbf{v}_4|\mathbf{0}) + C_R^{\perp_s}, (\mathbf{0}|\mathbf{v}_4) + C_R^{\perp_s}\}$ as a basis of $C_S^{\perp_s}/C_R^{\perp_s}$.

For a given secret $(m_1, m_2) \in \mathbf{F}_3^2$, the proposed encoder chooses a vector uniformly at random from the set

$$(0, 0, 0, m_1|0, 0, 0, m_2) + C_R^{\perp_s} \subset \mathbf{F}_3^8.$$

Since $|C_R^{\perp_s}| = 3^6$, for fixed $(m_1, m_2)$ the number of possible choices is $3^6$. But since $|\varphi\rangle$ is an eigenvector of all unitary matrices corresponding to a vector in $C_{\max}$, for fixed $(m_1, m_2)$ the number of possible quantum states is $|C_R^{\perp_s}/C_{\max}| = 3^2$. The encoded shares $X(\mathbf{a})Z(\mathbf{b})|\varphi\rangle$ consist of 4 qudits in $\mathbf{C}_3$. Each quantum share in $\mathbf{C}_3$ is distributed to each participant.

## 3 Necessary and sufficient conditions on qualified and forbidden sets

Let $A \subset \{1, ..., n\}$. Define $\mathbf{F}_p^A = \{(a_1, ..., a_n|b_1, ..., b_n) \in \mathbf{F}_p^{2n} : (a_i, b_i) = 0 \text{ for } i \notin A\}$. Let $P_A$ to be the projection map onto $A$, that is, $P_A(a_1, ..., a_n|b_1, ..., b_n) = (a_i|b_i)_{i \in A}$.

**Theorem 4** *For the secret sharing scheme described in Sect. 2, A is qualified if and only if*

$$\dim C_R/C_S = \dim C_R \cap \mathbf{F}_p^A/C_S \cap \mathbf{F}_p^A. \tag{1}$$

*A is forbidden if and only if*

$$0 = \dim C_R \cap \mathbf{F}_p^A/C_S \cap \mathbf{F}_p^A. \tag{2}$$

**Remark 5** The encoding procedure depends on the choice of $C_{\max} = C_{\max}^{\perp_s}$ but by Theorem 4 we see that the access structure is independent of that choice.

**Proof** (Theorem 4) Assume Eq. (1). Then there exists a basis $\{(\mathbf{a}_1|\mathbf{b}_1) + C_S, ..., (\mathbf{a}_k|\mathbf{b}_k) + C_S\}$ of $C_R/C_S$ such that $(\mathbf{a}_i|\mathbf{b}_i) \in \mathbf{F}_p^A$. Any two vectors in a coset $V \in C_S^{\perp_s}/C_R^{\perp_s}$ have the same value of the symplectic inner product against a fixed $(\mathbf{a}_i|\mathbf{b}_i)$, which will be denoted by $\langle(\mathbf{a}_i|\mathbf{b}_i), V\rangle_s$. Suppose that we have two different cosets $V_1, V_2 \in C_S^{\perp_s}/C_R^{\perp_s}$, and that $\langle(\mathbf{a}_i|\mathbf{b}_i), V_1\rangle_s = \langle(\mathbf{a}_i|\mathbf{b}_i), V_2\rangle_s$ for all $i$. It means that $V_1 - V_2 = C_R^{\perp_s}$ is zero in $C_S^{\perp_s}/C_R^{\perp_s}$, a contradiction. We have seen that any two different cosets have different symplectic inner product values against some $(\mathbf{a}_i|\mathbf{b}_i)$. For each $i$, the $n$ participants can collectively perform a quantum projective measurement corresponding to the eigenspaces of $X(\mathbf{a}_i)Z(\mathbf{b}_i)$ and can determine the symplectic inner product[1] $\langle(\mathbf{a}_i|\mathbf{b}_i), f(\mathbf{m})\rangle_s$ as [26, Lemma 5] when the classical secret is $\mathbf{m}$. Since $(\mathbf{a}_i|\mathbf{b}_i)$ has nonzero components only at $A$, the above measurement can be done only by $A$, which means $A$ can reconstruct $\mathbf{m}$.

Assume that Eq. (1) is false. Since the orthogonal space of $C_S$ in $\mathbf{F}_p^A$ is isomorphic to $P_A(C_S^{\perp_s})$, which is reminiscent of the duality between shortened linear codes and punctured linear codes [47], we see that $\dim P_A(C_S^{\perp_s})/P_A(C_R^{\perp_s}) < \dim C_S^{\perp_s}/C_R^{\perp_s}$. This means that there exist two different classical secrets $\mathbf{m}_1$ and $\mathbf{m}_2$ such that $P_A(f(\mathbf{m}_1)) = P_A(f(\mathbf{m}_2))$. This means that the encoding procedures of $\mathbf{m}_1$ and $\mathbf{m}_2$ are exactly the same on $A$ and produce the same density matrix on $A$, which shows that $A$ is not qualified.

---

[1] If we assume a non-prime finite field $\mathbf{F}_q$ as our base field, then the quantum measurement outcome just determines [26, Lemma 5] $\mathrm{Tr}_{q/p}(\langle(\mathbf{a}_i|\mathbf{b}_i), f(\mathbf{m})\rangle_s)$ in place of $\langle(\mathbf{a}_i|\mathbf{b}_i), f(\mathbf{m})\rangle_s$, where $\mathrm{Tr}_{q/p}$ is the trace map from $\mathbf{F}_q$ to its prime subfield $\mathbf{F}_p$. Assuming a non-prime field $\mathbf{F}_q$ significantly complicates the proofs of Theorem 4 and Lemma 8. So we assume a prime finite field until Remark 14.

Assume Eq. (2). Then we have dim $P_A(C_S^{\perp_s})/P_A(C_R^{\perp_s}) = 0$. This means that for all classical secrets $\mathbf{m}$, $P_A(f(\mathbf{m}))$ and their encoding procedures on $A$ are the same, which produces the same density matrix on $A$ regardless of $\mathbf{m}$. This shows that $A$ is forbidden.

Assume that Eq. (2) is false. Then there exist two different classical secrets $\mathbf{m}_1$, $\mathbf{m}_2$, and $(\mathbf{a}|\mathbf{b}) \in C_R \cap \mathbf{F}_p^A \setminus C_S \cap \mathbf{F}_p^A$ such that

$$\langle(\mathbf{a}|\mathbf{b}), f(\mathbf{m}_1)\rangle_s \neq \langle(\mathbf{a}|\mathbf{b}), f(\mathbf{m}_2)\rangle_s.$$

By [26, Lemma 5], this means that the quantum measurement corresponding to $X(\mathbf{a})Z(\mathbf{b})$ gives different outcomes with $\mathrm{Tr}_{\overline{A}}(\rho(\mathbf{m}_1))$ and $\mathrm{Tr}_{\overline{A}}(\rho(\mathbf{m}_2))$. Since $(\mathbf{a}|\mathbf{b}) \in \mathbf{F}_p^A$, measurement of $X(\mathbf{a})Z(\mathbf{b})$ can be performed only by participants in $A$. These observations show that $A$ is not forbidden. □

Next we give sufficient conditions in terms of the coset distance [16] or the first relative generalized Hamming weight [30]. To do so, we have to slightly modify them. For $(\mathbf{a}|\mathbf{b}) = (a_1, ..., a_n|b_1, ..., b_n) \in \mathbf{F}_p^n$, define its symplectic weight $\mathrm{swt}(\mathbf{a}|\mathbf{b}) = |\{i : (a_i, b_i) \neq (0, 0)\}|$. For $V_2 \subset V_1 \subset \mathbf{F}_p^{2n}$, we define their coset distance as $d_s(V_1, V_2) = \min\{\mathrm{swt}(\mathbf{a}|\mathbf{b}) : (\mathbf{a}|\mathbf{b}) \in V_1 \setminus V_2\}$.

**Theorem 6** *If* $|A| \leq d_s(C_R, C_S) - 1$ *then* $A$ *is forbidden. If* $|A| \geq n - d_s(C_S^{\perp_s}, C_R^{\perp_s}) + 1$ *then* $A$ *is qualified.*

**Example 7** Notations remain the same as Example 3. We have $d_s(C_R, C_S) = 3$ and $d_s(C_S^{\perp_s}, C_R^{\perp_s}) = 1$. By Theorem 6, we know that two or less participants are forbidden and all the participants are qualified.

**Proof** (Theorem 6) If $|A| \leq d_s(C_R, C_S) - 1$ then there is no $(\mathbf{a}|\mathbf{b}) \in C_R \cap \mathbf{F}_p^A \setminus C_S \cap \mathbf{F}_p^A$ and Eq. (2) holds.

Assume that $|A| \geq n - d_s(C_S^{\perp_s}, C_R^{\perp_s}) + 1$, or equivalently, $|\overline{A}| \leq d_s(C_S^{\perp_s}, C_R^{\perp_s}) - 1$. We have $C_S^{\perp_s} \cap \mathbf{F}_p^{\overline{A}} = C_R^{\perp_s} \cap \mathbf{F}_p^{\overline{A}}$. We also have $\mathbf{F}_p^{\overline{A}} = \ker(P_A)$, which means dim $P_A(C_S^{\perp_s}) -$ dim $P_A(C_R^{\perp_s}) = $ dim $C_S^{\perp_s} - $ dim $C_R^{\perp_s} = k$. Since dim $C_R \cap \mathbf{F}_p^A - $ dim $C_S \cap \mathbf{F}_p^A = $ dim $P_A(C_S^{\perp_s}) - $ dim $P_A(C_R^{\perp_s}) = k$, we see that Eq. (1) holds with $A$. □

## 4 Amount of information possessed by an intermediate set

Let $A \subset \{1, ..., n\}$ with $A \neq \emptyset$ and $A \neq \{1, ..., n\}$. In this section we study the amount of information possessed by $A$.

Because the result $f(\mathbf{m})$ of mapping $f$ is an element in $C_S^{\perp_s}/C_R^{\perp_s}$, any two vectors $(\mathbf{a}_1|\mathbf{b}_1)$ and $(\mathbf{a}_2|\mathbf{b}_2) \in f(\mathbf{m})$ give the same symplectic inner product values with any $(\mathbf{a}_3|\mathbf{b}_3) \in C_R$.

**Lemma 8** *For two classical secrets* $\mathbf{m}_1$ *and* $\mathbf{m}_2$, *we have*

- $\mathrm{Tr}_{\overline{A}}(\rho(\mathbf{m}_1)) = \mathrm{Tr}_{\overline{A}}(\rho(\mathbf{m}_2))$ *if and only if* $f(\mathbf{m}_1)$ *and* $f(\mathbf{m}_2)$ *give the same symplectic inner product for all vectors in* $C_R \cap \mathbf{F}_p^A$, *and*
- $\mathrm{col}(\mathrm{Tr}_{\overline{A}}(\rho(\mathbf{m}_1)))$ *and* $\mathrm{col}(\mathrm{Tr}_{\overline{A}}(\rho(\mathbf{m}_2)))$ *are orthogonal to each other if and only if* $f(\mathbf{m}_1)$ *and* $f(\mathbf{m}_2)$ *give different symplectic inner products for some vector* $(\mathbf{a}|\mathbf{b})$ *in* $C_R \cap \mathbf{F}_p^A$.

**Proof** Assume that $f(\mathbf{m}_1)$ and $f(\mathbf{m}_2)$ give the same symplectic inner product for all vectors in $C_R \cap \mathbf{F}_p^A$. Then we have $\{P_A(\mathbf{a}|\mathbf{b}) + P_A(C_R^{\perp_s}) : (\mathbf{a}|\mathbf{b}) + C_R^{\perp_s} \in f(\mathbf{m}_1)\} = \{P_A(\mathbf{a}|\mathbf{b}) + P_A(C_R^{\perp_s}) : (\mathbf{a}|\mathbf{b}) + C_R^{\perp_s} \in f(\mathbf{m}_2)\}$, and the encoding procedure on $A$ is the same for $\mathbf{m}_1$ and $\mathbf{m}_2$, which shows $\mathrm{Tr}_{\overline{A}}(\rho(\mathbf{m}_1)) = \mathrm{Tr}_{\overline{A}}(\rho(\mathbf{m}_2))$.

Assume that $f(\mathbf{m}_1)$ and $f(\mathbf{m}_2)$ give different symplectic inner product values for some vector $(\mathbf{a}|\mathbf{b})$ in $C_R \cap \mathbf{F}_p^A$. Then the quantum measurement corresponding to $X(\mathbf{a})Z(\mathbf{b})$ can be performed only by the participants in $A$ and by [26, Lemma 5] the outcomes for $\rho(\mathbf{m}_1)$ and $\rho(f(\mathbf{m}_2))$ are different with probability 1. This means that $\mathrm{col}(\mathrm{Tr}_{\overline{A}}(\rho(\mathbf{m}_1)))$ and $\mathrm{col}(\mathrm{Tr}_{\overline{A}}(\rho(\mathbf{m}_2)))$ are orthogonal to each other. □

**Proposition 9** *If* $\dim C_R \cap \mathbf{F}_p^A / C_S \cap \mathbf{F}_p^A = \ell$, *then the number of density matrices in* $\Lambda = \{\mathrm{Tr}_{\overline{A}}(\rho(\mathbf{m})) : \mathbf{m} \in \mathbf{F}_p^k\}$ *is* $p^\ell$.

*For a fixed density matrix* $\rho \in \Lambda$, *the number of classical secrets* $\mathbf{m}$ *such that* $\rho = \mathrm{Tr}_{\overline{A}}(\rho(\mathbf{m}))$ *is exactly* $p^{k-\ell}$.

*Proof* If $P_A(\mathbf{u}_1|\mathbf{v}_1) + P_A(C_R^{\perp_s}) \neq P_A(\mathbf{u}_2|\mathbf{v}_2) + P_A(C_R^{\perp_s})$ for $(\mathbf{u}_i|\mathbf{v}_i) \in f(\mathbf{m}_i)$ with classical secrets $\mathbf{m}_i$ $(i = 1, 2)$, then by Lemma 8 $\mathrm{col}(\mathrm{Tr}_{\overline{A}}(\rho(\mathbf{m}_1)))$ and $\mathrm{col}(\mathrm{Tr}_{\overline{A}}(\rho(\mathbf{m}_2)))$ are orthogonal. By the assumption, we have $\dim C_R \cap \mathbf{F}_p^A / C_S \cap \mathbf{F}_p^A = \dim P_A(C_S^{\perp_s}) / P_A(C_R^{\perp_s}) = \ell$. There are $p^\ell$ elements in $P_A(C^{\perp_s}) / P_A(C_{\max})$, which shows the first claim.

The composite $\mathbf{F}_p$-linear map "mod $P_A(C_R^{\perp_s})$" $\circ P_A \circ f$ from $\mathbf{F}_p^k$ to $P_A(C_S^{\perp_s}) / P_A(C_R^{\perp_s})$ is surjective. Thus the dimension of its kernel is $k - \ell$, which shows the second claim. □

**Definition 10** In light of Proposition 9, the amount of information possessed by a set $A$ of participants is defined as

$$(\log_2 p) \times \dim C_R \cap \mathbf{F}_p^A / C_S \cap \mathbf{F}_p^A = (\log_2 p) \times \dim P_A(C_S^{\perp_s}) / P_A(C_R^{\perp_s}). \quad (3)$$

**Remark 11** When the probability distribution of classical secrets $\mathbf{m}$ is uniform, the quantity in Definition 10 is equal to the Holevo information [45, Sect. 12.1.1] between $\mathbf{m}$ and $\mathrm{Tr}_{\overline{A}}(\rho(\mathbf{m}))$ by the same reason as [40, Remark 14].

We say that a secret sharing scheme is $r_i$-reconstructible if $|A| \geq r_i$ implies $A$ has $i \log_2 p$ or more bits of information [17]. We say that a secret sharing scheme is $t_i$-private if $|A| \leq t_i$ implies $A$ has less than $i \log_2 p$ bits of information [17]. In order to express $r_i$ and $t_i$ in terms of combinatorial properties of $C$, we review a slightly modified version of the relative generalized Hamming weight [30].

**Definition 12** [40] For two linear spaces $V_2 \subset V_1 \subset \mathbf{F}_p^{2n}$ and $i = 1, \ldots, k$, define the $i$-th relative generalized symplectic weight

$$d_s^i(V_1, V_2) = \min\{|A| : \dim \mathbf{F}_p^A \cap V_1 - \dim \mathbf{F}_p^A \cap V_2 \geq i\}. \quad (4)$$

Note that $d_s^1 = d_s$. The following theorem generalizes Theorem 6.

**Theorem 13**

$$t_i \geq d_s^i(C_R, C_S) - 1,$$
$$r_{k+1-i} \leq n - d_s^i(C_S^{\perp_s}, C_R^{\perp_s}) + 1.$$

*Proof* The following proof is almost the same as [40, Theorem 16]. Assume that $|A| \leq t_i$. By definition of $d_s^i$, $\dim C_R \cap \mathbf{F}_p^A / C_S \cap \mathbf{F}_p^A \leq i - 1$, which shows the first claim.

Assume that $|A| \geq r_i$. Then $|\overline{A}| \leq d_s^i(C_S^{\perp_s}, C_R^{\perp_s}) - 1$, which implies $\dim C_S^{\perp_s} \cap \mathbf{F}_p^{\overline{A}} / C_R^{\perp_s} \cap \mathbf{F}_p^{\overline{A}} \leq i - 1$. The last inequality implies $\dim C_R \cap \mathbf{F}_p^A / C_S \cap \mathbf{F}_p^A \geq k - i + 1$, which shows the second claim. □

**Remark 14** Here we have considered only the case of prime fields. Theorems 4, 6, 13, Proposition 9 and Definition 10 can be generalized to arbitrary finite fields, similarly to what has been done in [40, Sect. 5.1].

## 5 Gilbert–Varshamov-type existential condition

Let $q$ be some prime power. In this section, we give a sufficient condition (5) for existence of $C_S \subset C_R \subseteq C_R^{\perp_s} \subset C_S^{\perp_s} \subset \mathbf{F}_q^{2n}$, with given parameters.

**Theorem 15** *If positive integers $n$, $k$, $s$, $\delta_t$, $\delta_r$ satisfy*

$$\frac{q^{n+k+s} - q^{n+s}}{q^{2n}-1} \sum_{i=1}^{\delta_r-1} \binom{n}{i}(q^2-1)^i + \frac{q^{n-s}-q^{n-k-s}}{q^{2n}-1} \sum_{i=1}^{\delta_t-1} \binom{n}{i}(q^2-1)^i < 1, \quad (5)$$

*then there exist $C_S$ and $C_R$ such that $C_S \subset C_R \subseteq C_R^{\perp_s} \subset C_S^{\perp_s} \subset \mathbf{F}_q^{2n}$, $\dim C_S = n - k - s$, $\dim C_R = n - s$ $d_s(C_S^{\perp_s}, C_R^{\perp_s}) \geq \delta_r$ and $d_s(C_R, C_S) \geq \delta_t$.*

**Proof** The following argument is similar to the proof of Gilbert-Varshamov bound for stabilizer codes [9] and also to [40]. Let $\mathrm{Sp}(q, n)$ be the set of symplectic isometries on $\mathbf{F}_q^{2n}$, that is, bijective linear maps preserving the values of the symplectic inner product. Let $A(k)$ be the set of pairs of linear spaces $(V, W)$ such that $\dim V = n - k - s$, $\dim W = n - s$ and $V \subset W \subseteq W^{\perp_s} \subset V^{\perp_s} \subset \mathbf{F}_q^{2n}$. For $\mathbf{e} \in \mathbf{F}_q^{2n}$, define $B_V(k, \mathbf{e}) = \{(V, W) \in A(k) : \mathbf{e} \in V^{\perp_s} \setminus W^{\perp_s}\}$ and $B_W(k, \mathbf{e}) = \{(V, W) \in A(k) : \mathbf{e} \in W \setminus V\}$.

For nonzero $\mathbf{e}_1$, $\mathbf{e}_2 \in \mathbf{F}_q^{2n}$, we have $|B_W(k, \mathbf{e}_1)| = |B_W(k, \mathbf{e}_2)|$, whose proof is the same argument as [40, Proof of Theorem 25], and reproduced below: For nonzero $\mathbf{e}_1$, $\mathbf{e}_2 \in \mathbf{F}_q^{2n}$ with $M_1\mathbf{e}_1 = \mathbf{e}_2$ ($M_1 \in \mathrm{Sp}(q, n)$) and some fixed $(V_1, W_1) \in A(k)$, we have

$$\begin{aligned}
&|B_W(k, \mathbf{e}_1)| \\
&= |\{(v, W) \in A(k) : \mathbf{e}_1 \in W \setminus V\}| \\
&= |\{(MV_1, MW_1) : \mathbf{e}_1 \in MW \setminus MV, M \in \mathrm{Sp}(q, n)\}| \\
&= |\{(M_1^{-1}MV_1, M_1^{-1}MW_1) : \mathbf{e}_1 \in M_1^{-1}MW \setminus M_1^{-1}MV, M \in \mathrm{Sp}(q, n)\}| \\
&= |\{(MV_1, MW_1) : M_1\mathbf{e}_1 \in MW \setminus MV, M \in \mathrm{Sp}(q, n)\}| \\
&= |\{(MV_1, MW_1) : \mathbf{e}_2 \in MW \setminus MV, M \in \mathrm{Sp}(q, n)\}| \\
&= |\{(V, W) \in A(k) : \mathbf{e}_2 \in W \setminus V\}| \\
&= |B_W(k, \mathbf{e}_2)|.
\end{aligned}$$

By a similar argument we also have $|B_V(k, \mathbf{e}_1)| = |B_V(k, \mathbf{e}_2)|$.

For each $(V, W) \in A(k)$, the number of $\mathbf{e}$ such that $\mathbf{e} \in W \setminus V$ is $|W| - |V| = q^{n-s} - q^{n-k-s}$. The number of triples $(\mathbf{e}, V, W)$ such that $\mathbf{0} \neq \mathbf{e} \in W \setminus V$ is

$$\sum_{\mathbf{0} \neq \mathbf{e} \in \mathbf{F}_q^{2n}} |B_W(k, \mathbf{e})| = |A(k)| \times (q^n - q^k),$$

which implies

$$\frac{|B_W(k, \mathbf{e})|}{|A(k)|} = \frac{q^{n-s} - q^{n-k-s}}{q^{2n}-1}. \quad (6)$$

Similarly we have

$$\frac{|B_V(k, \mathbf{e})|}{|A(k)|} = \frac{q^{n+k+s} - q^{n+s}}{q^{2n}-1}. \quad (7)$$

If there exists $(V, W) \in A(k)$ such that $(V, W) \notin B_V(k, \mathbf{e}_1)$ and $(V, W) \notin B_V(k, \mathbf{e}_2)$ for all $1 \leq \mathrm{swt}(\mathbf{e}_1) \leq \delta_r - 1$ and $1 \leq \mathrm{swt}(\mathbf{e}_2) \leq \delta_t - 1$ then there exists a pair of $(V, W)$ with the desired properties. The number of $\mathbf{e}$ such that $1 \leq \mathrm{swt}(\mathbf{e}) \leq \delta - 1$ is given by

$$\sum_{i=1}^{\delta-1} \binom{n}{i} (q^2 - 1)^i. \tag{8}$$

By combining Eqs. (6), (7) and (8) we see that Eq. (5) is a sufficient condition for ensuring the existence of $(V, W)$ required in Theorem 15. □

We will derive an asymptotic form of Theorem 15.

**Theorem 16** *Let $R \leq 1$, $S \leq 1$, $\epsilon_t < 0.5$ and $\epsilon_r < 0.5$ be nonnegative real numbers. Define $h_q(x) = -x \log_q x - (1-x) \log_q (1-x)$. For sufficiently large $n$, if*

$$h_q(\epsilon_t) + \epsilon_t \log_q(q^2 - 1) < 1 + S \text{ and}$$
$$h_q(\epsilon_r) + \epsilon_r \log_q(q^2 - 1) < 1 - R - S,$$

*then there exist $C_S$ and $C_R$ such that $C_S \subset C_R \subseteq C_R^{\perp_s} \subset C_S^{\perp_s} \subset \mathbf{F}_q^{2n}$, $\dim C_S = n - \lfloor n(R + S) \rfloor$, $\dim C_R = n - \lfloor nS \rfloor$ $d_s(C_S^{\perp_s}, C_R^{\perp_s}) \geq \lfloor n\epsilon_r \rfloor$ and $d_s(C_R, C_S) \geq \lfloor n\epsilon_t \rfloor$.*

***Proof*** Proof can be done by almost the same argument as [43, Sect. III.C], reproduced as below.

$$\sum_{i=1}^{\delta-1} \binom{n}{i} (q^2 - 1)^i$$
$$\leq (\delta - 1) \binom{n}{\delta - 1} (q^2 - 1)^{\delta-1}$$
$$= \exp_q \left[ \log_q(\delta - 1) + n h_q \left( \frac{\delta - 1}{n} \right) + n \frac{\delta - 1}{n} \log_q(q^2 - 1) \right]. \tag{9}$$

Note that we can find $\binom{n}{i} \leq \exp_q[h_q(i/n)]$ for $i < n/2$ in [31].

When the assumption of Theorem 16 holds, by Eq. (9) we see that $\log_q[$ left hand side of Eq. (5) $]/n$ goes to a negative value, noting that $\lim_{n \to \infty} \frac{\log_q(\delta-1)}{n} = 0$. This concludes the proof. □

In [40, Theorem 26] we proved a special case $S = 0$ of Theorem 16. The new parameter $S \geq 0$ provides larger flexibility.

# 6 Strong security

Let $n = q$, and let $k$, $n - s - k$ be nonnegative even integers. The field size $q$ can be either odd or even. We will consider the case that the number of participants is smaller than $q$ in Remark 21. Let $\alpha_1, ..., \alpha_n \in \mathbf{F}_q$ be $n$ distinct elements. Define an $[n, k]$ Reed-Solomon (RS) code as

$$\mathrm{RS}(n, k) = \{(g(\alpha_1), \ldots, g(\alpha_n)) : g(x) \in \mathbf{F}_q[x], \deg g(x) < k\}.$$

Then $\mathrm{RS}(n, k)^{\perp_E} = \mathrm{RS}(n, n - k)$ because $n = q$.

## 6.1 Insecure example

In order to justify our study of strong security, we will show an insecure ramp scheme constructed in the framework of [38,40]. Assume that $n = q$ are even integers only in Sect. 6.1. Let $C_S = \{\mathbf{0}\}$, $s = 0$, $k = n$, and $C_R = C_{\max} = C_R^{\perp s} = RS(n, n/2) \times RS(n, n/2)$. For classical secret $\mathbf{m} = (m_1, \ldots, m_n)$, let $h_1(x) = m_1 x^{n/2} + \cdots + m_{n/2} x^{n-1}$ and $h_2(x) = m_{1+n/2} x^{n/2} + \cdots + m_n x^{n-1}$. Define an $\mathbf{F}_q$-linear map $f : \mathbf{F}_q^n \to C_S^{\perp s}/C_R^{\perp s}$ in Sect. 2.2 as

$$f(\mathbf{m}) = (h_1(\alpha_1), \ldots, h_1(\alpha_n)|h_2(\alpha_1), \ldots, h_2(\alpha_n)) + C_R^{\perp s}.$$

As shown in [38,40, Sect. 5.4], any $n-1$ shares have $(n-2) \log_2 q$ bits of information about $\mathbf{m}$. Assume $\alpha_n = 0$. The participant set $A = \{1, \ldots, n-1\}$ can perform a measurement corresponding to a nonzero vector in $C_R \cap \mathbf{F}_q^A$, which contains $(\mathbf{u}^i|\mathbf{0})$ and $(\mathbf{0}|\mathbf{u}^i)$ for $i = 1$, $\ldots, n/2 - 1$, where $\mathbf{u} = (\alpha_1^i, \ldots, \alpha_{n-1}^i, \alpha_n^i = 0)$. We have

$$\langle (\mathbf{u}^i|\mathbf{0}), (h_1(\alpha_1), \ldots, h_1(\alpha_n)|h_2(\alpha_1), \ldots, h_2(\alpha_n)) \rangle_s = m_{n-i},$$
$$\langle (\mathbf{0}|\mathbf{u}^i), (h_1(\alpha_1), \ldots, h_1(\alpha_n)|h_2(\alpha_1), \ldots, h_2(\alpha_n)) \rangle_s = -m_{n/2-i-1}.$$

By [26, Lemma 5], the share set $A = \{1, \ldots, n-1\}$ can completely determine $n-2$ symbols $m_1, \ldots, m_{n/2-2}, m_{n/2}, \ldots, m_{n/2-1}$ in the classical secret $\mathbf{m}$. In the next subsection, we will show a remedy to address this kind of insecurity.

## 6.2 Definition and construction of strongly secure schemes

**Definition 17** Let $A \subset \{1, \ldots, n\}$ be a share set and $\rho_A$ the density matrix of shares in $A$. Let $\mathbf{m} \in \mathbf{F}_q^k$ be a classical secret drawn from the uniform probability distribution on $\mathbf{F}_q^k$. Let $Z \subset \{1, \ldots, k/2\}$. A quantum ramp secret sharing scheme is said to be *strongly secure* if $I(\mathbf{m}; \rho_A) = \ell \log_2 q > 0$ then $I(P_{Z \cup k/2+Z}(\mathbf{m}); \rho_A) = 0$ for all $Z$ with $2|Z| \leq k - \ell$, where $I(\cdot; \cdot)$ denotes the Holevo information [45, Section 12.1.1.1] counted in $\log_2$, $k/2 + Z = \{k/2 + z : z \in Z\}$ and $P_{Z \cup k/2+Z}$ is previously defined projection to an index set $Z \cup k/2 + Z \subset \{1, \ldots, k\}$.

The above definition is a straightforward generalization of [24, Definition 6] to the quantum setting, with regarding $(m_i, m_{i+k/2}) \in \mathbf{F}_q^2$ as one symbol and the secret $\mathbf{m}$ consisting of $k/2$ such symbols.

In this subsection, we will construct a scheme distributing a classical secret consisting of $k$ symbols in $\mathbf{F}_q$ to $n$ participants with 1 qudit of dimension $q$, so that any $(n+k+s)/2$ participants can reconstruct the secret, while any $(n+s)/2$ or less participants have no information about the secret, with the above strong security. We note that a somewhat similar idea was used for construction of a strongly secure ramp secret sharing scheme with classical shares [39].

We assume that $\alpha_1, \ldots, \alpha_{k/2}$ are nonzero. Define

$$C_S = \{(\mathbf{a}|\mathbf{b}) : \mathbf{a}, \mathbf{b} \in RS(n, (n-k-s)/2)\},$$
$$C_R = \{(\mathbf{a}|\mathbf{b}) : \mathbf{a}, \mathbf{b} \in RS(n, (n-s)/2)\}.$$

Then we can easily see that

$$C_R^{\perp s} = \{(\mathbf{a}|\mathbf{b}) : \mathbf{a}, \mathbf{b} \in RS(n, (n+s)/2)\},$$
$$C_S^{\perp s} = \{(\mathbf{a}|\mathbf{b}) : \mathbf{a}, \mathbf{b} \in RS(n, (n+k+s)/2)\},$$
$$\dim C_S = n - k - s,$$
$$\dim C_R = n - s.$$

We can choose $C_{\max}$ as, for example,

$$C_{\max} = \{(\mathbf{a}|\mathbf{b}) : \mathbf{a} \in RS(n, \lfloor n/2 \rfloor), \mathbf{b} \in RS(n, \lceil n/2 \rceil)\}.$$

For a classical secret $\mathbf{m} = (m_1, \ldots, m_k) \in \mathbf{F}_q^k$, find $g_1(x) = a_0 x^0 + \cdots + a_{k/2-1} x^{k/2-1}$ and $g_2(x) = b_0 x^0 + \cdots + b_{k/2-1} x^{k/2-1}$ such that $g_1(\alpha_j) = m_j/\alpha_j^{(n+s)/2}$ and $g_2(\alpha_j) = m_{j+k/2}/\alpha_j^{(n+s)/2}$ for all $j = 1, \ldots, k/2$. Such $g_1(x)$ and $g_2(x)$ always exist because computation of $g_i(x)$ is just the inverse mapping of the encoding of $RS(k/2, k/2)$ for the codeword $(m_1/\alpha_1^{(n+s)/2}, \ldots, m_{k/2}/\alpha_{k/2}^{(n+s)/2})$. Let $g_3(x) = x^{(n+s)/2} g_1(x)$ and $g_4(x) = x^{(n+s)/2} g_2(x)$. Observe that $g_3(\alpha_j) = m_j$ and $g_4(\alpha_j) = m_{k/2+j}$. Define a bijective $\mathbf{F}_q$-linear map $f$ as

$$f(\mathbf{m}) = (g_3(\alpha_1), \ldots, g_3(\alpha_n)|g_4(\alpha_1), \ldots, g_4(\alpha_n)) + C_R^{\perp s} \in C_S^{\perp s}/C_R^{\perp s}.$$

The quantum shares are computed as in Sect. 2.2 with the above $f$. For $A \subset \{1, \ldots, n\}$, let $\rho_A$ be the density matrix of quantum shares in $A$. By almost the same argument as [40, Sect. 5.4], we see that the Holevo information $I(\mathbf{m}; \rho_A)$ between $\mathbf{m}$ and $\rho_A$ is

$$I(\mathbf{m}; \rho_A) = \begin{cases} 0 & \text{if } 0 \leq |A| \leq \frac{n+s}{2}, \\ 2\left(|A| - \frac{n+s}{2}\right) \log_2 q & \text{if } \frac{n+s}{2} \leq |A| \leq \frac{n+k+s}{2}, \\ k \log_2 q & \text{if } \frac{n+k+s}{2} \leq |A| \leq n. \end{cases} \tag{10}$$

In particular, the above means that $A$ is qualified if and only if $|A| \geq (n+k+s)/2$ and $A$ is forbidden if and only if $|A| \leq (n+s)/2$.

Let $B \subset \{1, \ldots, k\}$. By slight abuse of notation, by $P_B(\mathbf{m})$ we mean $(m_i)_{i \in B}$. In order to verify the strong security, we have to compute the Holevo information $I(P_B(\mathbf{m}); \rho_A)$. In order to compute $I(P_B(\mathbf{m}); \rho_A)$, we consider the following related problem. Let $\overline{B} = \{1, \ldots, k\} \setminus B$. When we consider the strong security of $P_B(\mathbf{m})$, the rest $P_{\overline{B}}(\mathbf{m})$ serves as dummy variable to hide $P_B(\mathbf{m})$.

Let $B' \subset \{1, \ldots, k/2\}$ and $\overline{B'} = \{1, \ldots, k/2\} \setminus B'$. For $g(x) = a_0 x^0 + \cdots a_{(n+k+s)/2-1-|B'|} x^{(n+k+s)/2-1-|B'|}$, define $g_{B'}(x) = a_{(n+k+s)/2-|B'|} x^{(n+k+s)/2-|B'|} + \cdots + a_{(n+k+s)/2-1} x^{(n+k+s)/2-1}$ such that $g_{B'}(\alpha_j) = -\sum_{i=(n+s)/2}^{(n+k+s)/2-1-|B'|} a_i \alpha_j^i$ for $j \in B'$. Such a $g_{B'}(x)$ is uniquely determined because it is the inverse of encoding of $[|B'|, |B'|]$ generalized Reed-Solomon code. Define a linear code

$$D_{B'} = \{(g(\alpha_1) + g_{B'}(\alpha_1), \ldots, g(\alpha_n) + g_{B'}(\alpha_n)) : \deg g(x) \leq (n+k+s)/2 - 1 - |B'|\}.$$

For a subset $S \subset \mathbf{F}_q^n$, by abuse of notation we mean $P_A(S) = \{(x_i)_{i \in A} : (x_1, \ldots, x_n) \in S\}$.

**Lemma 18** *Assume* $|B'| < k/2$.

$$\dim P_A(RS(n, (n+k+s)/2)) - \dim P_A(D_{B'})$$
$$= \begin{cases} 0 & \text{if } 0 \leq |A| \leq \frac{n+k+s}{2} - |B'|, \\ \left(|A| + |B'| - \frac{n+k+s}{2}\right) & \text{if } \frac{n+k+s}{2} - |B'| \leq |A| \leq \frac{n+k+s}{2}, \\ |B'| & \text{if } \frac{n+k+s}{2} \leq |A| \leq n. \end{cases} \tag{11}$$

**Proof** Since the minimum Hamming distance of $\mathrm{RS}(n, (n + k + s)/2)$ is $(n - k - s)/2 + 1$, we have [47]

$$\dim P_A(\mathrm{RS}(n, (n + k + s)/2)) = \begin{cases} |A| & \text{if } 0 \le |A| \le \frac{n+k+s}{2}, \\ \frac{n+k+s}{2} & \text{if } \frac{n+k+s}{2} \le |A| \le n. \end{cases} \quad (12)$$

The codeword in $D_{B'}$ is the sum of a codeword in $\mathrm{RS}(n, (n + k + s)/2 - |B'|)$ and the codeword defined by $g_{B'}(x)$. The latter can be seen as a codeword in a generalized Reed-Solomon code of length $n$ and dimension $|B'|$. So, the Hamming weight of a codeword defined by $g_{B'}(x)$ is $\ge n + 1 - |B'|$. There exists a codeword in $\mathrm{RS}(n, (n + k + s)/2 - |B'|)$ of Hamming weight $(n - k - s)/2 + 1 + |B'|$. Since $|B'| < k/2$, the condition $k - s \le n$ implies $n + 1 - |B'| > (n - k - s)/2 + 1 + |B'|$. Under this condition, the minimum weight codeword in $\mathrm{RS}(n, (n + k + s)/2 - |B'|)$ cannot be canceled by a codeword defined by $g_{B'}(x)$. Therefore, the minimum Hamming distance of $D_{B'}$ is $(n - k - s)/2 + 1 + |B'|$, which, by [47], implies

$$\dim P_A(D_{B'}) = \begin{cases} |A| & \text{if } 0 \le |A| \le (n + k + s)/2 - |B'|, \\ (n + k + s)/2 - |B'| & \text{if } (n + k + s)/2 - |B'| \le |A| \le n. \end{cases} \quad (13)$$

Combining Eqs. (12) and (13) gives the claim of this lemma. $\qquad\square$

In light of Eq. (11), define $\ell(a, b)$ as

$$\ell(a, b) = \begin{cases} 0 & \text{if } 0 \le a \le \frac{n+k+s}{2} - b, \\ \left(a + b - \frac{n+k+s}{2}\right) & \text{if } \frac{n+k+s}{2} - b \le a \le \frac{n+k+s}{2}, \\ b & \text{if } \frac{n+k+s}{2} \le a \le n. \end{cases}$$

**Proposition 19** *Let* $B_1 = B \cap \{1, \ldots, k/2\}$ *and* $B_2 = B \cap \{1 + k/2, \ldots, k\}$

$$I(P_B(\mathbf{m}); \rho_A) = [\ell(|A|, |B_1|) + \ell(|A|, |B_2|)] \log_2 q$$

**Proof** Let $\overline{B} = \{1, \ldots, k\} \setminus B$. When we see $P_B(\mathbf{m})$ as secret and $P_{\overline{B}}(\mathbf{m})$ as meaningless dummy randomness, the corresponding secret sharing scheme is described by $C_{\mathrm{S}}^{\perp s} \supset C_{\mathrm{R}}'^{\perp s} \supset C_{\mathrm{R}}' \supset C_{\mathrm{S}}$, where $C_{\mathrm{R}}'^{\perp s}$ corresponds to $C_{\mathrm{R}}^{\perp s}$ in Sect. 2.2, and

$$C_{\mathrm{R}}'^{\perp s} = \{(\mathbf{a}|\mathbf{b}) : \mathbf{a} \in D_{B_1}, \mathbf{b} \in D_{B_2}\}$$

In order to evaluate $I(P_B(\mathbf{m}); \rho_A)$, we have to compute $\dim C_{\mathrm{R}}' \cap \mathbf{F}_q^A / C_{\mathrm{S}} \cap \mathbf{F}_q^A$, which is equal to $\dim P_A(C_{\mathrm{S}}^{\perp s}) - \dim P_A(C_{\mathrm{R}}'^{\perp s})$. By Lemma 18 we have

$$\dim P_A(C_{\mathrm{S}}^{\perp s}) - \dim P_A(C_{\mathrm{R}}'^{\perp s}) = \ell(|A|, |B_1|) + \ell(|A|, |B_2|),$$

which completes the proof. $\qquad\square$

**Corollary 20** *The proposed encoding scheme is strongly secure in the sense of Definition 17.*

**Proof** Assume $I(\mathbf{m}; \rho_A) = \ell \log_2 q > 0$. Then, by Eq. (10) $|A| = \frac{\ell + n + s}{2}$, or equivalently, $\ell = 2|A| - n - s$. Assume $2|Z| \le k - \ell$. Then $|A| \le \frac{n+k+s}{2} - |Z|$. Application of Proposition 19 with $|B_1| = |B_2| = |Z|$ shows $I(P_{Z \cup k/2 + Z}(\mathbf{m}); \rho_A) = 0$. $\qquad\square$

**Remark 21** Although we have assumed $n = q$, we note that the number $n'$ of participants can be made smaller than $q$ by discarding shares, whose access structure is the same as

$$C_S = \mathbf{F}_q^A \cap \{(\mathbf{a}|\mathbf{b}) : \mathbf{a}, \mathbf{b} \in RS(n, (n - k - s)/2)\},$$

$$C_R = \mathbf{F}_q^A \cap \{(\mathbf{a}|\mathbf{b}) : \mathbf{a}, \mathbf{b} \in RS(n, (n - s)/2)\},$$

$$C_R^{\perp_s} = P_A(\{(\mathbf{a}|\mathbf{b}) : \mathbf{a}, \mathbf{b} \in RS(n, (n + s)/2)\}),$$

$$C_S^{\perp_s} = P_A(\{(\mathbf{a}|\mathbf{b}) : \mathbf{a}, \mathbf{b} \in RS(n, (n + k + s)/2)\}),$$

where $A \subset \{1, \ldots, n\}$ with $|A| = n'$.

### 6.3 Comparison with the McEliece–Sarwate scheme

McEliece and Sarwate [44] proposed the first strongly secure ramp secret sharing scheme, whose strong security was proved much later [46]. Let $\alpha_1, \ldots, \alpha_{n+k}$ be distinct elements in $\mathbf{F}_q$. For a given secret $(m_1, \ldots, m_k) \in \mathbf{F}_q^k$, randomly choose a polynomial $g(x)$ of degree less than $(n + k + s)/2$ such that $g(\alpha_i) = m_i$ for $i = 1, \ldots, k$. Then it distributes $g(\alpha_{k+i})$ to the $i$-th participant. Any $(n + k + s)/2$ or more participants can reconstruct the secret. Any $(n - k + s)/2$ or less participants have no information about the secret. Thus the qualified sets are the same, but the McEliece–Sarwate scheme has smaller forbidden sets than the proposed one in Sect. 6.2. Equivalently, the classical secret in the proposed construction can be twice as large as the McEliece-Sarwate scheme for the same qualified sets and the same forbidden sets. In addition, the McEliece-Sarwate scheme can support at most $q - k$ participants, while the proposed one in Sect. 6.2 can support at most $q$ participants.

**Acknowledgements** The author would like to thank anonymous reviewers for careful reading and detailed comments.

### Compliance with ethical standards